\documentclass[12pt]{article}
\usepackage{enumerate}
\usepackage{amsfonts}
\usepackage{amsmath}
\usepackage{amssymb}
\usepackage{amsthm}

\def\H{\mathcal{H}}
\def\K{\mathcal{K}}
\def\P{\mathcal{P}}
\def\S{\mathfrak{S}}
\def\T{\mathfrak{T}}
\def\B{\mathfrak{B}}

\newcommand{\id}{\mathrm{Id}}
\newcommand{\Tr}{\mathrm{Tr}}

\newcounter{defin}  \newcounter{lemma}  \newcounter{theorem}
\newcounter{property} \newcounter{corol}  \newcounter{remark} \newcounter{example}

\newenvironment{lemma}{\par\refstepcounter{lemma}
     \textbf{Lemma \thelemma.} }{\rm\par}
\newenvironment{theorem}{\par\refstepcounter{theorem}
     \textbf{Theorem \thetheorem.}\ }{\rm\par}
\newenvironment{property}{\par\refstepcounter{property}
     \textbf{Proposition \theproperty.}\ }{\rm\par}
\newenvironment{corollary}{\par\refstepcounter{corol}
     \textbf{Corollary \thecorol.} }{\rm\par}
\newenvironment{definition}{\par\refstepcounter{defin}
     \textbf{Definition \thedefin.}\ }{\rm\par}
\newenvironment{remark}{\par\refstepcounter{remark}
     \textbf{Remark \theremark.}}{\rm\par}

\begin{document}
\title{Monotonicity of the Holevo quantity: a necessary condition for equality in terms of a channel and its applications.}
\author{M.E. Shirokov\footnote{email:msh@mi.ras.ru}\\
Steklov Mathematical Institute, RAS, Moscow}
\date{}
\maketitle 
\begin{abstract}
A condition for reversibility
(sufficiency) of a channel with respect to a given countable family of states with bounded rank is obtained.

This condition shows that a quantum channel preserving the Holevo quantity
of at least one (discrete or continuous) ensemble of states with
rank  $\leq r$ has the r-partially entanglement-breaking
complementary channel. Several applications of this result are
considered. In particular, it is shown that coincidence of the
constrained Holevo capacity and the quantum mutual information of a
quantum channel at least at one full rank state implies that this
channel is entanglement-breaking.
\end{abstract}
\vspace{15pt}
\tableofcontents

\section{Introduction}

The Holevo quantity $\chi(\{\pi_i,\rho_i\})$ of an ensemble of
quantum states $\{\pi_i,\rho_i\}$ provides an upper bound for
accessible classical information which can be obtained by applying a
quantum measurement \cite{H-b}. The fundamental monotonicity property of
the relative entropy implies non-increasing of the Holevo quantity
under action of an arbitrary quantum channel $\Phi$, that is
\begin{equation}\label{chi-q-m+}
\chi(\{\pi_i,\Phi(\rho_i)\})\leq\chi(\{\pi_i,\rho_i\})
\end{equation}
for any  ensemble of quantum states $\{\pi_i,\rho_i\}$. \medskip

Necessary and sufficient conditions for the case of equality in
fundamental entropic inequalities of quantum theory have been
intensively studied (see \cite{Petz,J&R,Petz+,Ruskai,Z&W} and the references
therein). In particular, two characterizations of the equality in
(\ref{chi-q-m+}) in finite dimensions are obtained in \cite[Examples
4 and 9]{Petz}. The first one derived from  Petz's theorem (Theorem
\ref{petz-t} in Appendix 6.1) states that the equality in
(\ref{chi-q-m+}) holds if and only if
\begin{equation}\label{inv-cond}
\rho_i=A\Phi^*(B\,\Phi(\rho_i)B)A,\quad A=(\bar{\rho})^{1/2},\;
B=(\Phi(\bar{\rho}))^{-1/2},\quad \forall i,
\end{equation}
where $\Phi^*$ is a dual map to the channel $\Phi$ and $\bar{\rho}$
is the average state of the ensemble $\{\pi_i,\rho_i\}$. The second
characterization of the equality in (\ref{chi-q-m+}) is derived from
the characterization of the equality case in the strong
subadditivity of the quantum entropy by identifying the channel
$\Phi$ with a subchannel of a partial trace, so it is not clear how
to apply this condition to a given quantum channel $\Phi$.\smallskip

Condition (\ref{inv-cond}) means reversibility
(sufficiency) of the channel $\Phi$ with respect to the set $\{\rho_i\}$ of quantum states \cite{J&P,J-rev,P-sqc}. \smallskip

In Section 3 we prove a simple necessary
condition for reversibility
of a channel with respect to a given countable family of states with bounded rank, which implies
a necessary condition for the equality in (\ref{chi-q-m+}) expressed in terms of
the channel $\Phi$.  The main advantage of this condition consists
in possibility to use it in analysis of entropic characteristics of
a given quantum channel determined as extremal values of particular
functionals depending on the Holevo quantity (such as the Holevo
capacity and the related characteristics).

In Section 4 we generalize the above condition to the case of
continuous ensembles.

Several applications of the obtained conditions concerning the
notions of the Holevo capacity and of the minimal output entropy of
a quantum channel as well as properties of the quantum conditional
entropy are considered in Section 5. In particular, it is shown that
the equality in the general inequality
$$
\bar{C}(\Phi,\rho)\leq I(\Phi,\rho),
$$
connecting the constrained Holevo capacity $\bar{C}(\Phi,\rho)$ and the quantum mutual
information $I(\Phi,\rho)$ of a quantum channel $\Phi$ at a state $\rho$, implies
that the restriction of the channel $\Phi$ to the set of states
supported by the subspace $\,\mathrm{supp}\rho\,$ is
entanglement-breaking.

\section{Preliminaries} Let $\H,\K $ be either finite dimensional or
separable Hilbert spaces, $\B(\H)$ and
$\mathfrak{T}( \mathcal{H})$ -- the Banach spaces of all bounded
operators in $\mathcal{H}$ and of all trace-class operators in
$\H$  correspondingly, $\B_+(\H)$ -- the positive cone in $\B(\H)$ and $\S(\H)$ -- the closed convex subset
of $\mathfrak{T}( \H)$ consisting of positive operators
with unit trace called \emph{states} \cite{B&R,N&Ch}.

Denote by $I_{\mathcal{H}}$ and $\mathrm{Id}_{\mathcal{H}}$ the
unit operator in a Hilbert space $\mathcal{H}$ and the identity
transformation of the Banach space $\mathfrak{T}(\mathcal{H})$
correspondingly.

A linear completely positive trace preserving map
$\Phi:\mathfrak{T}(\mathcal{H}_A)\rightarrow\mathfrak{T}(\mathcal{H}_B)$
is called  \emph{quantum channel} \cite{N&Ch}. We will say that the above  channel
$\Phi$ is \emph{isometrically equivalent} to the channel
$\,\Phi':\mathfrak{T}(\mathcal{H}_A)\rightarrow\mathfrak{T}(\mathcal{H}_B')\,$ if
there is a partial isometry
$W:\mathcal{H}_B\rightarrow\mathcal{H}_{B'}$ such that
\begin{equation}\label{c-isom}
\Phi'(A)=W\Phi(A)W^*,\quad\Phi(A)=W^*\Phi'(A)W,\quad
A\in \T(\H_A).
\end{equation}

For a given channel $\Phi:\mathfrak{T}(\mathcal{H}_A)\rightarrow\mathfrak{T}(\mathcal{H}_B)$ the Stinespring
theorem implies existence of a Hilbert space $\mathcal{H}_E$ and of
an isometry
$V:\mathcal{H}_A\rightarrow\mathcal{H}_B\otimes\mathcal{H}_E$ such
that
\begin{equation}\label{Stinespring-rep}
\Phi(A)=\mathrm{Tr}_{\mathcal{H}_E}VA V^{*},\quad
A\in\mathfrak{T}(\mathcal{H}_A).
\end{equation}
A quantum  channel
\begin{equation}\label{c-channel}
\mathfrak{T}(\mathcal{H}_A)\ni
A\mapsto\widehat{\Phi}(A)=\mathrm{Tr}_{\mathcal{H}_B}VAV^{*}\in\mathfrak{T}(\mathcal{H}_E)
\end{equation}
is called \emph{complementary} to the channel $\Phi$
\cite{H-c-c}.\footnote{The quantum channel $\widehat{\Phi}$ is
also called \emph{conjugate} to the channel $\Phi$ \cite{KMNR}.} The
complementary channel is defined uniquely in the following sense: if
$\widehat{\Phi}':\mathfrak{T}(\mathcal{H}_A)\rightarrow\mathfrak{T}(\mathcal{H}_{E'})$
is a channel defined by (\ref{c-channel}) via the Stinespring isometry
$V':\mathcal{H}_A\rightarrow\mathcal{H}_B\otimes\mathcal{H}_{E'}$ then
the channels $\widehat{\Phi}$ and $\widehat{\Phi}'$ are isometrically equivalent in the sense of (\ref{c-isom}) \cite{H-c-c}.

The Stinespring representation (\ref{Stinespring-rep}) is called
\emph{minimal} if the subspace
$$
\mathcal{M}=\left\{\,(X\otimes
I_E)V|\varphi\rangle\;|\;\varphi\in\H_A,\, X\in\B(\H_B)\,\right\}
$$
is dense in $\mathcal{H}_B\otimes\mathcal{H}_E$. The complementary
channel $\widehat{\Phi}$ defined by (\ref{c-channel}) via the
minimal Stinespring representation has the following property:
\begin{equation}\label{full-rank}
\widehat{\Phi}(\rho)\;\text{is a full rank state
in}\;\S(\H_E)\;\text{for any full rank
state}\;\rho\;\textrm{in}\;\S(\H_A).
\end{equation}

The Stinespring representation (\ref{Stinespring-rep}) generates the
Kraus representation
\begin{equation}\label{Kraus-rep}
\Phi(A)=\sum_{k}V_{k}AV^{*}_{k},\quad A\in\,\mathfrak{T}(\mathcal{H}),
\end{equation}
where $\{V_{k}\}$ is the set of bounded linear
operators from $\mathcal{H}_A$ into $\mathcal{H}_B$ such that
$\sum_{k}V^{*}_{k}V_{k}=I_{\H_A}$ defined by the
relation
$$
\langle\varphi|V_k\psi\rangle=\langle\varphi\otimes
k|V\psi\rangle,\quad\varphi\in\H_B,\psi\in\H_A,
$$
where $\{|k\rangle\}$ is a particular orthonormal basis in the
space $\mathcal{H}_E$. The corresponding complementary channel is
expressed as follows
\begin{equation}\label{Kraus-rep-c}
\widehat{\Phi}(A)=\sum_{k,l}\mathrm{Tr}\left[V_{k}AV_{l}^{*}\right]|k\rangle\langle
l|,\quad A\in\,\mathfrak{T}(\mathcal{H}).
\end{equation}


The Schmidt rank of a pure state $\omega$ in
$\S(\H\otimes\K)$ can be defined as the operator rank of the isomorphic
states $\Tr_{\K}\omega$ and $\Tr_{\H}\omega$ \cite{T&H}.

The Schmidt class $\,\S_r$ of order $r\in\mathbb{N}\,$ is
the minimal convex closed subset of $\S(\H\otimes\K)$ containing all
pure states with the Schmidt rank $\leq r$, i.e. $\S_r$ is the
convex closure of these pure states
\cite{T&H,Sh-18}.\footnote{In finite dimensions the convex closure
coincides with the convex hull by the Caratheodory theorem, but in infinite dimensions even the set of all \emph{countable} convex mixtures of pure states with the Schmidt rank $\leq r$ is a proper subset of $\S_r$ for each $r$ \cite{Sh-18}.} In this
notation $\S_1$ is the set of all separable (non-entangled) states
in $\S(\H\otimes\K)$.

A channel $\Phi$ is called \emph{entanglement-breaking} if for an arbitrary
Hilbert space $\K$ the state $\Phi\otimes\id_{\K}(\omega)$ is
separable for any state $\omega\in\S(\H_A\otimes\K)$ \cite{e-b-ch}.
This notion is generalized in \cite{p-e-b-ch} as follows.\medskip

\begin{definition}\label{p-e-b-ch-d}
A channel $\Phi:\T(\H_A)\rightarrow\T(\H_B)$ is called
$r$-\emph{partially entanglement-breaking }(briefly $r$-PEB) if for
an arbitrary Hilbert space $\K$  the state
$\Phi\otimes\id_{\K}(\omega)$ belongs to the Schmidt class $\S_r\subset\S(\H_B\otimes\K)$
for any state $\omega\in\S(\H_A\otimes\K)$.\medskip
\end{definition}

In this notation entanglement-breaking channels are $1$-PEB
channels. Properties of $r$-PEB channels in finite dimensions are
studied in \cite{p-e-b-ch}, where it is proved, in particular, that
the class of $r$-PEB channels coincides with the class of channels
having  Kraus representation (\ref{Kraus-rep}) such that
$\mathrm{rank}V_k\leq r$ for all $k$. But in infinite dimensions the
first class is essentially wider than the second one, moreover, for
each $r$ there exist $r$-PEB channels  such that all operators in
any their Kraus representations have infinite rank
\cite{Sh-18}.\medskip

If a channel $\Phi$ has Kraus representation (\ref{Kraus-rep}) such that
$\mathrm{rank}V_k=1$ for all $k$ then representation (\ref{Kraus-rep-c}) shows that
the complementary channel $\widehat{\Phi}$ is
pseudo-diagonal in the sense of the following definition
\cite{R&Co,H-c-c,KMNR}.\smallskip
\begin{definition}\label{p-d-ch}
A channel $\Phi:\S(\H_A)\rightarrow\S(\H_B)$ is called
\emph{pseudo-diagonal} if it has the representation
$$
\Phi(\rho)=\sum_{i,j}c_{ij}\langle
\psi_i|\rho|\psi_j\rangle|i\rangle\langle j|,\quad\rho\in\S(\H_A),
$$
where $\|c_{ij}\|$ is a Gram matrix of some collection of unit vectors,
$\{|\psi_i\rangle\}$ is a collection of vectors in $\H_A$ satisfying
the overcompleteness relation $\;\sum_i |\psi_i\rangle\langle
\psi_i|=I_{\H_A}\,$  and $\{|i\rangle\}$ is an orthonormal basis in
$\H_B$.
\end{definition}\medskip
\pagebreak

Let $H(\rho)$ and $H(\rho\|\sigma)$ be respectively the von Neumann
entropy of the state $\rho$ and the quantum relative entropy of the
states $\rho$ and $\sigma$ \cite{L,N&Ch,O&P}.\medskip

A finite or countable collection of states $\{\rho_i\}$ with the
corresponding probability distribution $\{\pi_i\}$ is called
\emph{ensemble} and denoted $\{\pi_i,\rho_i\}$. The state
$\bar{\rho}=\sum_i \pi_i\rho_i$ is called the \emph{average state}
of the ensemble $\{\pi_i,\rho_i\}$.\medskip

The Holevo quantity of an ensemble $\{\pi_i,\rho_i\}$ is defined as
follows
\begin{equation*}
\chi(\{\pi_i,\rho_i\})\doteq\sum_i\pi_i
H(\rho_i\|\bar{\rho})=H(\bar{\rho})-\sum_i\pi_i H(\rho_i),
\end{equation*}
where the second expression is valid under the condition
$H(\bar{\rho})<+\infty$.

By monotonicity of the relative entropy for an arbitrary quantum
channel $\Phi$ we have
\begin{equation}\label{chi-q-m}
\chi(\{\pi_i,\Phi(\rho_i)\})\leq\chi(\{\pi_i,\rho_i\}).
\end{equation}

\begin{remark}\label{ent-g-c}
If $H(\bar{\rho})<+\infty$ and $H(\Phi(\bar{\rho}))<+\infty$ then
inequality (\ref{chi-q-m}) means convexity of the entropy gain
$H(\Phi(\rho))-H(\rho)$ of the channel $\Phi$.
\end{remark}
\medskip

A necessary condition for the equality in (\ref{chi-q-m}) expressed in terms of the channel $\Phi$ is obtained in the next section (Corollary \ref{main-th-c}).

\section{A condition for reversibility of a channel with respect to a countable set of states}

Let $\{\rho_i\}$ be a  finite or countable set of states in $\S(\H)$
and $\{\pi_i\}$ be a non-generate probability distribution. By Petz's theorem (Theorem \ref{petz-t} in the Appendix 6.1)
if the Holevo quantity of an ensemble $\{\pi_i,\rho_i\}$ is finite then the equality in (\ref{chi-q-m}) holds if and only if the channel $\Phi$ is \emph{reversible} with respect to the set $\{\rho_i\}$ in the sense of the following definition. \medskip

\begin{definition}\label{rev-def} \cite{J-rev}
A channel $\Phi:\S(\H_A)\rightarrow\S(\H_B)$ is reversible with respect to a set $\S\subseteq\S(\H_A)$ if there exists a channel
$\,\Psi:\S(\H_B)\rightarrow\S(\H_A)$ such that
$\,\rho=\Psi\circ\Phi(\rho)\,$ for all $\,\rho\in\S$.\footnote{This property is also called sufficiency of
the channel $\Phi$ with respect to the set $\S$ \cite{J&P,P-sqc}.}
\end{definition}\medskip

The following theorem gives a necessary condition for reversibility
of a channel with respect to a countable set of states with bounded rank.\medskip

\begin{theorem}\label{main} \emph{Let $\Phi:\S(\H_A)\rightarrow\S(\H_B)$ be a quantum
channel and \break $\widehat{\Phi}:\S(\H_A)\rightarrow\S(\H_E)$ be
its complementary channel. Let $\{\rho_i\}^n_{i=1},$ $n\leq+\infty$,
be a set of states in $\,\S(\H_A)$ such that $\,\sup_i\Tr A\rho_i>0$ for any nonzero operator $A$ in $\B_{+}(\H_A)$
and $\,\mathrm{rank}\rho_i\leq r\in\mathbb{N}\,$ for all $\,i$.}

\emph{If the channel $\Phi$ is reversible with  respect to the set
$\,\{\rho_i\}$ then the channel $\,\widehat{\Phi}$ has Kraus
representation (\ref{Kraus-rep}) such that $\;\mathrm{rank}V_k\leq
r$  for all $\,k$ and hence $\,\widehat{\Phi}$ is a $r$-partially
entanglement-breaking channel (Def.\ref{p-e-b-ch-d}).}
\smallskip

\emph{If the above hypothesis holds with $\;r=1\;$, i.e.
$\rho_i=|\varphi_i\rangle\langle\varphi_i|$ for all $i$, then the
channel $\,\Phi$ is isometrically equivalent \footnote{By Lemma \ref{isom-eq} below the reversibility with respect to a given set of states is a common property for two isometrically equivalent channels.} (in the sense of (\ref{c-isom})) to
the pseudo-diagonal channel
\begin{equation}\label{ch-rep}
\Phi'(\rho)=\!\!\sum_{i,\,j,\,k,\,l}\langle\phi_i|\rho|\phi_k\rangle
\langle\psi_{kl}|\psi_{ij}\rangle|i\otimes j\rangle\langle k\otimes
l|
\end{equation}
from $\S(\H_A)$ into $\S(\H_n\otimes\H_B)$, where $\{|\phi_i\rangle\}^n_{i=1}$ is an overcomplete system of
vectors in $\H_A$ defined by means of an arbitrary non-generate
probability distribution $\{\pi_i\}^n_{i=1}$ as follows
\begin{equation}\label{v-phi-rep}
|\phi_i\rangle=\pi_i^{1/2}(\bar{\rho}_{\pi})^{-1/2}|\varphi_i\rangle,\quad
\bar{\rho}_{\pi}=\sum^n_{i=1}\pi_i|\varphi_i\rangle\langle\varphi_i|,
\end{equation}
$\{|\psi_{ij}\rangle\}$ is a collection of vectors such that
$\sum_j\|\psi_{ij}\|^2=1$ for all $\,i$, $\{|i\rangle\}^n_{i=1}$ and
$\{|j\rangle\}$ are orthonormal base in $\H_n$ and in $\H_B$
correspondingly.}
\end{theorem}
\medskip

The main assertion of Theorem \ref{main} means that the channel
$\widehat{\Phi}$ has the following property:  for an arbitrary
Hilbert space $\K$  and any state $\omega$ in $\S(\H_A\otimes\K)$ the
state $\widehat{\Phi}\otimes\id_{\K}(\omega)$ is a \emph{countably
decomposable} state in the Schmidt class $\S_r\subset\S(\H_E\otimes\K)$, i.e. it can be
represented as a countable convex mixture of pure states having the
Schmidt rank $\leq r$  (there exist states in $\S_r$ which are not countably
decomposable \cite{Sh-18}).

The last assertion of Theorem \ref{main} gives a \emph{necessary and sufficient} condition for reversibility of the channel
$\Phi$ provided the set $\{|\varphi_i\rangle\langle\varphi_i|\}$ consists of orthogonal states (in this case $|\phi_i\rangle=|\varphi_i\rangle$ for all $i$ and $\H_n=\H_A$).

\textbf{Proof.}  Let
$\,\widehat{\Phi}(\rho)=\sum_{k=1}^{m}V_k\rho V_k^*$,
$\,m\leq+\infty$, be the Kraus representation of the channel
$\widehat{\Phi}:\S(\H_A)\rightarrow\S(\H_E)$ obtained via its
minimal Stinespring representation with the isometry
$V:\H_A\rightarrow\H_E\otimes\H_{C}$ (see Section 2). The
complementary channel $\Psi=\widehat{\widehat{\Phi}}$ to the channel
$\widehat{\Phi}$ defined via this representation is expressed as
follows
$$
\S(\H_A)\ni\rho\mapsto\Psi(\rho)=\sum_{k,l=1}^m \Tr V_k\rho
V_l^*|k\rangle\langle l|\in\S(\H_{C}),
$$
where $\,\{|\,k\rangle\}_{k=1}^m$ is an orthonormal basis in the
$m$-dimensional Hilbert space $\H_C$.

Since $\Psi=\widehat{\widehat{\Phi}}$, there exists a partial
isometry $W:\H_B\rightarrow\H_C$ such that
$$
\Psi(\rho)=W\Phi(\rho)W^*,\quad\Phi(\rho)=W^*\Psi(\rho)W,\quad
\rho\in \S(\H_A).
$$
By Lemma \ref{isom-eq} below the channel $\Psi$ is reversible with respect to the set $\{\rho_i\}$.\smallskip

Let $\{\pi_i\}^n_{i=1}$ be an arbitrary non-generate probability
distribution and $\bar{\rho}$ be the average state of the
ensemble $\{\pi_i,\rho_i\}^n_{i=1}$. By property (\ref{full-rank})
$\Psi(\bar{\rho})$ is a full rank state in $\S(\H_C)$. By Theorem 3 in \cite{J&P} the reversibility condition implies $A_i=\Psi^*(B_i)$ for all $i$, where
$A_i=\pi_i(\bar{\rho})^{-1/2}\rho_i(\bar{\rho})^{-1/2}$ and
$B_i=\pi_i(\Psi(\bar{\rho}))^{-1/2}\Psi(\rho_i)(\Psi(\bar{\rho}))^{-1/2}$
are positive operators in $\B(\H_A)$ and in $\B(\H_{C})$
correspondingly.

Note that
$$
\Psi^*(A)=\sum_{k,l=1}^m \langle l |A|k\rangle V_l^*V_k,\quad
A\in\B(\H_C).
$$

Let $B_i=\sum_j|\psi_{ij}\rangle\langle \psi_{ij}|$, where
$\,\{|\psi_{ij}\rangle\}_j$ is a set of vectors in $\H_C$, for each
$i$.\footnote{This representation can be obtained by multiplying the
both sides of the equality $I_{\H_C}=\sum_j|j\rangle\langle j|$,
where $\,\{|j\rangle\}$ is an arbitrary basis in $\H_C$, by
$B_i^{1/2}$.} Since $\Psi(\bar{\rho})$ is a full rank state in
$\S(\H_C)$, we have
\begin{equation*}
\sum_{i,j}|\psi_{ij}\rangle\langle \psi_{ij}|=\sum_i B_i=I_{\H_C}.
\end{equation*}
By Lemma \ref{new-kraus-rep} below
$\,\widehat{\Phi}(\rho)=\sum_{i,j}W_{ij}\rho W_{ij}^*$, where
$W_{ij}=\sum^m_{k=1}\langle\psi_{ij}|k\rangle V_k$.\smallskip

Since $A_i=\Psi^*(\sum_j|\psi_{ij}\rangle\langle \psi_{ij}|)\,$ is
an operator of rank $\leq r$ for each $i$ and
$$
\Psi^*(|\psi_{ij}\rangle\langle \psi_{ij}|)=\sum_{k,l=1}^m \langle l
|\psi_{ij}\rangle\langle\psi_{ij}|k\rangle
V_l^*V_k=\,W_{ij}^*W_{ij},
$$
the family $\{W_{ij}\}$ consists of operators of rank $\leq r$.
\smallskip

If $\rho_i=|\varphi_i\rangle\langle\varphi_i|$ then   $A_i=|\phi_{i}\rangle\langle\phi_{i}|$, where the vector
$|\phi_{i}\rangle$ is defined by (\ref{v-phi-rep}). Hence
representation (\ref{ch-rep}) can be obtained from the above
arguments by using representation (\ref{Kraus-rep-c}) for the channel complementary to the channel $\,\widehat{\Phi}(\rho)=\sum_{i,j}W_{ij}\rho W_{ij}^*$ and by noting that the above partial isometry $W^*$ is an embedding of $\H_C$ into $\H_B$ (since $\Psi(\bar{\rho})$ is a full rank state in $\S(\H_C)$). $\square$
\smallskip

\begin{lemma}\label{isom-eq}
\emph{Let $\,\Phi:\S(\H_A)\rightarrow\S(\H_B)$ and
$\,\Phi':\S(\H_A)\rightarrow\S(\H_{B'})$ be quantum channels isometrically
equivalent in the sense of (\ref{c-isom}). If the channel $\,\Phi$
is reversible with respect to a set $\,\S\subseteq\S(\H_A)$ then the
channel $\,\Phi'$ is reversible with respect to the set $\,\S$ and
vice versa.}
\end{lemma}\smallskip

\textbf{Proof.} Let $\Psi$ be the reverse channel for the channel
$\Phi$, i.e. $\Psi\circ\Phi(\rho)=\rho$ for all $\rho\in\S$.
Let $\Theta(\cdot)=W^*(\cdot)W+\sigma\Tr(I_{\H_{B'}}-WW^*)(\cdot)$ be a channel from $\S(\H_{B'})$ into $\S(\H_{B})$, where
$W$ is the partial isometry from (\ref{c-isom}) and $\sigma$ is a fixed state in $\S(\H_{B})$.  Then $\Psi\circ\Theta$ is a reverse channel for the
channel $\Phi'$. $\square$
\smallskip

\begin{lemma}\label{new-kraus-rep}
\emph{Let $\,\Phi(\rho)=\sum_{k=1}^{m}V_k\rho V_k^*$ be a quantum
channel and $\,\{|k\rangle\}_{k=1}^m$ be an orthonormal basis in the
$m$-dimensional Hilbert space $\H_m$, $m\leq+\infty$. An arbitrary
overcomplete system $\{|\psi_i\rangle\}$ of vectors in $\H_m$
generates the Kraus representation $\,\Phi(\rho)=\sum_{i}W_i\rho
W_i^*$ of the channel $\,\Phi$, where
$W_i=\sum^m_{k=1}\langle\psi_i|k\rangle V_k$.}
\end{lemma}\smallskip

\textbf{Proof.} Since $\sum_{i}|\psi_i\rangle\langle
\psi_i|=I_{\H_m}$, we have
$$
\begin{array}{c}
\displaystyle\sum_{i}W_i\rho W^*_i=\sum^m_{k,l=1} V_k\rho V^*_l
\sum_{i}\langle\psi_i|k\rangle\langle
l|\psi_i\rangle\\\displaystyle=\sum^m_{k,l=1} V_k\rho V^*_l
\sum_{i}\Tr |k\rangle\langle
l||\psi_i\rangle\langle\psi_i|=\sum^m_{k=1} V_k\rho V^*_k. \;\square
\end{array}
$$

By Petz's theorem (Theorem 3 in Appendix 6.1) Theorem \ref{main} implies the following necessary condition for the equality in (\ref{chi-q-m}), which is not sufficient (even in the weak sense) by Remark \ref{necessity} below.\smallskip

\begin{corollary}\label{main-th-c} \emph{Let $\Phi:\S(\H_A)\rightarrow\S(\H_B)$ be a quantum
channel and \break $\widehat{\Phi}:\S(\H_A)\rightarrow\S(\H_E)$ be
its complementary channel.} \emph{If there exists an ensemble
$\{\pi_i,\rho_i\}$ with the full rank average state $\bar{\rho}$ such that
$\,\mathrm{rank}\rho_i\leq r$ for all $\,i$ and
\begin{equation*}
\chi(\{\pi_i,\Phi(\rho_i)\})=\chi(\{\pi_i,\rho_i\})<+\infty
\end{equation*}
then the channel $\,\widehat{\Phi}$ has Kraus representation
(\ref{Kraus-rep}) such that $\,\mathrm{rank}V_k\leq r$  for all
$\,k$ and hence $\,\widehat{\Phi}$ is a $r$-partially
entanglement-breaking channel (Def.\ref{p-e-b-ch-d}).}
\end{corollary}\pagebreak

\begin{remark}\label{main-r}
By Corollary \ref{main-th-c} to prove the strict inequality in (\ref{chi-q-m})
for all ensembles $\{\pi_i,\rho_i\}$ such that $\mathrm{supp}\,\bar{\rho}=\H_A$ and $\mathrm{rank}\rho_i\leq r$ for all $i$ it suffices to show that the
channel $\widehat{\Phi}$ is not $r$-partially entanglement-breaking. This can be done by showing existence of a state $\omega$ in $\S(\H_A\otimes\K)$ such that
\begin{equation}\label{s-c}
\text{either}\quad
SN(\widehat{\Phi}\otimes\id_{\K}(\omega))>r\quad
\text{or}\quad E(\widehat{\Phi}\otimes\id_{\K}(\omega))>\log r,
\end{equation}
where $SN$ is the Schmidt number (defined in \cite{T&H} and in \cite{Sh-18} in finite and in infinite dimensions correspondingly) and $E$ is any convex entanglement monotone coinciding on the set of pure states with the entropy of a partial state, in particular, $E=EoF$ \cite{P&V}.

The condition $\,\mathrm{supp}\,\bar{\rho}=\H_A\,$ in Corollary \ref{main-th-c} can be removed by considering the restrictions of the channels $\Phi$ and $\widehat{\Phi}$ to the set $\S(\H_{\bar{\rho}})$, where $\H_{\bar{\rho}}=\mathrm{supp}\,\bar{\rho}$. Thus, to prove the strict inequality in (\ref{chi-q-m})
for an arbitrary ensemble $\{\pi_i,\rho_i\}$ such that $\mathrm{rank}\rho_i\leq r$  for all $i$ it suffices to show existence of a state $\omega$ in $\S(\H_{\bar{\rho}}\otimes\K)$ such  that (\ref{s-c}) holds.

The necessity of the condition $\,\mathrm{supp}\,\bar{\rho}=\H_A\,$ is discussed in Remark \ref{full-rank-n} in Section 5.1.
\medskip
\end{remark}

We complete this section by the following remark.\smallskip

\begin{remark}\label{necessity} There exist quantum channels
complementary to entanglement breaking channels such that the strict
inequality holds in (\ref{chi-q-m}) for any ensemble of pure states
with the full rank average. To show this consider the
channel
$$
\Phi(\rho)=\sum_{k=1}^3\langle\varphi_k|\rho|\varphi_k\rangle|k\rangle\langle
k|,
$$
where $|\varphi_{k}\rangle=\sqrt{\frac{2}{3}}\left[\cos\frac{2}{3}\pi (k-1), \sin\frac{2}{3}\pi (k-1)\right]^\mathrm{T}$, $k=1,2,3$,
are vectors in the 2-D space $\H_A$ and $\{|k\rangle\}_{k=1}^3$ is
an orthonormal basis in the 3-D space $\H_B$.

Suppose there exists an ensemble $\{\pi_i,\rho_i\}$ of pure states
with the full rank average state $\bar{\rho}$ such that
$\chi(\{\pi_i,\Phi(\rho_i)\})=\chi(\{\pi_i,\rho_i\})$.  Since
$\Phi(\bar{\rho})$ is a full rank state and
$\Phi^*(A)=\sum_{k=1}^3\langle k|A|k\rangle|\varphi_k\rangle\langle
\varphi_k|$, condition (\ref{inv-cond}) implies that
$\mathrm{rank}\,\Phi(\rho_i)=1$ for any $i$. But this can not be
valid, since it is easy to see that $\mathrm{rank}\,\Phi(\rho)>1$
for any $\rho$. Hence
$\chi(\{\pi_i,\Phi(\rho_i)\})<\chi(\{\pi_i,\rho_i\})$ for any
ensemble $\{\pi_i,\rho_i\}$ of pure states with the full rank
average. $\square$
\end{remark}

\section{Continuous ensembles}

A continuous (generalized) ensemble of quantum states can be defined
as a Borel probability measure $\mu$ on the set $\S(\H)$. The Holevo
quantity of such ensemble (measure) $\mu$ is defined as follows (cf.
\cite{H-Sh-2})
\begin{equation}\label{h-q-c}
\chi(\mu)=\int_{\mathfrak{S}(\mathcal{H})}H(\rho\Vert\bar{\rho}(\mu))\mu(d\rho),
\end{equation}
where $\bar{\rho}(\mu)$ is the barycenter of the measure $\mu$
defined by the Bochner integral
$$
\bar{\rho}(\mu)=\int_{\mathfrak{S}(\mathcal{H})}\rho \mu(d\rho ).
$$
If $H(\bar{\rho}(\mu))<+\infty$ then
$\chi(\mu)=H(\bar{\rho}(\mu))-\int_{\mathfrak{S}(\mathcal{H})}H(\rho)\mu
(d\rho)$ \cite{H-Sh-2}.\smallskip

Denote by $\P(\mathcal{A})$ the set of all Borel probability
measures on a closed subset $\mathcal{A}\subset\T(\H)$ endowed
with the weak convergence topology \cite{Par}.\smallskip

The image of a continuous ensemble $\mu\in\P(\S(\H_A))$ under a
channel $\Phi:\S(\H_A)\rightarrow\S(\H_B)$ is a continuous ensemble
corresponding to the measure
$\Phi(\mu)\doteq\mu\circ\Phi^{-1}\in\P(\S(\H_B))$. Its Holevo
quantity can be expressed as follows
\begin{equation}\label{chi-phi-mu}
\begin{array}{c}
\displaystyle\chi(\Phi(\mu))\doteq\int_{\mathfrak{S}(\H_A)}H(\Phi(\rho
)\Vert \Phi (\bar{\rho}(\mu)))\mu(d\rho )\\\displaystyle=H(\Phi
(\bar{\rho}(\mu)))-\int_{\mathfrak{S}(\H_A)}H(\Phi(\rho))\mu
(d\rho),
\end{array}
\end{equation}
where the second formula is valid under the condition $H(\Phi
(\bar{\rho}(\mu)))<+\infty$.\medskip

We will assume in what follows that $\bar{\rho}(\mu)$ is a full rank state in $\S(\H_A)$ and that $\sup_{\rho\in\S(\H_A)}\Tr B\Phi(\rho)>0$ for any $B\in\B_+(\H_B)\setminus\{0\}$ (otherwise we may consider restrictions to smaller subspaces $\H'_A\subset\H_A$ and $\H'_B\subset\H_B$). It follows from these assumptions that
$\Phi(\bar{\rho}(\mu))$ is a full rank state in $\S(\H_B)$.\medskip

Similarly to the discrete case monotonicity of the relative entropy
implies monotonicity of the Holevo quantity for continuous
ensembles:
\begin{equation}\label{chi-d++}
\chi(\Phi(\mu))\leq\chi(\mu).
\end{equation}
By using Petz's theorem (Theorem \ref{petz-t} in Appendix 6.1) one can obtain the following characterization of the equality in (\ref{chi-d++}).
\smallskip
\begin{property}\label{PJ-cont-ver}
\emph{Let $\,\Phi:\S(\H_A)\rightarrow\S(\H_B)$ be a quantum
channel and $\mu$ be a measure in $\P(\S(\H_A))$ such that $\,\chi(\mu)<+\infty$. Let $\,\Theta_{\bar{\rho}(\mu)}$ be the predual channel to the linear
completely positive unital map}
\begin{equation*}
\Theta^*_{\bar{\rho}(\mu)}(\cdot)=A\Phi\left(B(\cdot)B\right)A,\quad
A=[\Phi(\bar{\rho}(\mu))]^{-1/2},\; B=[\bar{\rho}(\mu)]^{1/2}.
\end{equation*}
\emph{The following statements are equivalent:}
\begin{enumerate}[(i)]
  \item $\chi(\Phi(\mu))=\chi(\mu)$;
  \item \emph{$H(\Phi(\rho)\Vert \Phi (\bar{\rho}(\mu)))=H(\rho
\Vert \bar{\rho}(\mu))$ for $\mu$-almost all $\rho$ in $\,\S(\H_A)$;}
  \item \emph{$\rho=\Theta_{\bar{\rho}(\mu)}(\Phi(\rho))$ for $\mu$-almost
all $\rho$ in $\,\S(\H_A)$;}
  \item \emph{the channel $\,\Phi$ is reversible with respect to $\mu$-almost
all $\rho$ in $\,\S(\H_A)$.}
\end{enumerate}
\end{property}

In contrast to Theorem 3 in \cite{J&P}, in Proposition
\ref{PJ-cont-ver} it is not assumed that the "dominating" state $\bar{\rho}(\mu)$ is a countable convex mixture of some states from the support of the measure $\mu$.
\medskip

Suppose the support $\S_{\mu}$ of the measure $\mu$ consists of states with rank $\leq r$. By Proposition
\ref{PJ-cont-ver} the equality in (\ref{chi-d++}) implies existence of a subset $\S\subseteq\S_{\mu}$
such that $\mu(\S)=1$ and $\rho=\Theta_{\bar{\rho}(\mu)}(\Phi(\rho))$ for all $\rho\in \S$. By Lemma 2  in
\cite{J&P} there exists an ensemble $\{\pi_i,\rho_i\}$ of states in
$\S$ having the average state $\bar{\rho}$ such that
$\mathrm{supp}\rho\subseteq\mathrm{supp}\bar{\rho}$ for all
$\rho\in\S$ and hence $\bar{\rho}$ is a full rank state in
$\S(\H_A)$ (since $\bar{\rho}(\mu)=\int_{\S}\rho\mu(d\rho)$
is a full rank state). By applying Theorem \ref{main} to the set $\{\rho_i\}$ we obtain the following continuous version (in fact, a
generalization) of Corollary \ref{main-th-c}. \medskip

\begin{theorem}\label{main++} \emph{Let $\,\Phi:\S(\H_A)\rightarrow\S(\H_B)$ be a quantum
channel and \break $\widehat{\Phi}:\S(\H_A)\rightarrow\S(\H_E)$ be
its complementary channel.} \emph{If there exists a measure
$\mu\in\P(\S^r)$, where
$\,\S^r=\{\rho\in\S(\H_A)\,|\,\mathrm{rank}\,\rho\leq r\}$, with the
full rank barycenter $\,\bar{\rho}(\mu)$ such that
\begin{equation}\label{chi-nd++}
\chi(\Phi(\mu))=\chi(\mu)<+\infty,
\end{equation}
then the channel $\,\widehat{\Phi}$ has Kraus representation
(\ref{Kraus-rep}) such that $\,\mathrm{rank}V_k\leq r$  for all
$\,k$ and hence $\,\widehat{\Phi}$ is a $r$-partially
entanglement-breaking channel (Def.\ref{p-e-b-ch-d}).}
\smallskip

\emph{If the above hypothesis holds with $\,r=1$ then the channel $\,\Phi$ is isometrically equivalent to a pseudo-diagonal channel (Def.\ref{p-d-ch}) in the sense of (\ref{c-isom}).}\medskip
\end{theorem}

\begin{remark}\label{main++r+}
Condition (\ref{chi-nd++}) in Theorem \ref{main++} can be
replaced by the condition of reversibility of the channel
$\Phi$ with respect to $\mu$-almost all
$\rho$ in $\S(\H_A)$, in which finiteness of $\chi(\mu)$ is not
required.
\end{remark}\medskip
\smallskip

\section{Applications}

\subsection{Finite dimensional channels}

In this subsection we  consider some implications of Corollary \ref{main-th-c} assuming that $\dim\H_A$ and $\dim\H_B$ are finite. \smallskip

The Holevo capacity of the channel $\Phi:\S(\H_A)\rightarrow\S(\H_B)$ is defined as follows (cf.\cite{N&Ch})
\begin{equation}\label{chi-cap}
\bar{C}(\Phi)=\sup_{\{\pi_i,\rho_i\}}\chi(\{\pi_i,\Phi(\rho_i)\}).
\end{equation}
Monotonicity the Holevo quantity shows that
\begin{equation*}
\bar{C}(\Phi)\leq\log\dim\H_A
\end{equation*}
for any channel $\,\Phi$. The equality holds in the above inequality
for many quantum channels (for example, for the noiseless channel, for the channel $\Phi(\rho)=\sum_{k}\langle k|\rho|k\rangle|k\rangle\langle k|$,
where $\{|k\rangle\}$ is an orthonormal basis in $\H_B=\H_A$).

Since the supremum in (\ref{chi-cap}) is always achieved at some ensembles
of pure states \cite{Sch-West}, Corollary \ref{main-th-c} with $r=1$ implies the following observation.
\medskip
\begin{property}\label{main-c-2}
\emph{Let $\,\Phi:\S(\H_A)\rightarrow\S(\H_B)$ be a quantum channel. If $\,\bar{C}(\Phi)=\log\dim\H_A$ then
the channel $\,\Phi$ is isometrically equivalent to a pseudo-diagonal channel
in the sense of (\ref{c-isom}) and hence it is degradable \cite{R&Co}.}
\end{property}
\medskip

Proposition \ref{main-c-2} can be used to show positivity of the
minimal output entropy
$$
H_{\mathrm{min}}(\Phi)=\min_{\rho\in\S(\H_A)}H(\Phi(\rho))
$$
for a  class of quantum channels.\medskip

\begin{corollary}\label{main-c-3}
\emph{Let $\,\Phi:\S(\H_A)\rightarrow\S(\H_B)$, $\H_B=\H_A$, be a
quantum channel covariant with respect to some irreducible
representation $\{V_g\}_{g\in G}$ of a compact group $G$ in the
sense that $\,\Phi(V_g\rho V^*_g)=V_g\Phi(\rho)V^*_g$ for all $g\in
G$. If the channel $\,\Phi$ is not isometrically equivalent to a pseudo-diagonal channel (in particular, is not degradable) then
$\,H_{\mathrm{min}}(\Phi)>0$.}
\end{corollary}\medskip

\textbf{Proof.} It follows from the covariance condition of the
corollary that $\bar{C}(\Phi)=\log\dim\H_A-H_{\mathrm{min}}(\Phi)$
\cite{H-r-c-c}. By Proposition \ref{main-c-2} we have
$H_{\mathrm{min}}(\Phi)>0$. $\square$
\medskip

Corollary \ref{main-c-3} shows that
$H_{\mathrm{min}}(\Phi)>0$ for any unital qubit channel, which
is not isometrically equivalent to a pseudo-diagonal channel (in particular, is not degradable).\medskip

\subsection{Infinite dimensional channels}

In this subsection we consider two implications of Theorems
\ref{main} and \ref{main++} concerning general (finite or infinite
dimensional) quantum systems and channels.

\subsubsection{Strict decrease of the Holevo quantity under partial trace and strict concavity of the conditional entropy}

Since the partial trace
$\S(\H\otimes\K)\ni\rho\mapsto\Tr_{\H}\rho\,$ is not $r$-PEB channel
for $\,r<\dim\K$, Corollary \ref{main-th-c} and Theorem \ref{main++} imply the
following observations.
\medskip

\begin{property}\label{main-c-4} \emph{Let $\,\H_A=\H_B\otimes\H_E$ and
$\;\Phi(\rho)=\Tr_{\H_E}\rho$, $\rho\in\S(\H_A)$.}\smallskip

A) \emph{$\chi(\{\pi_i,\Phi(\rho_i)\})<\chi(\{\pi_i,\rho_i\})$ for
any ensemble $\{\pi_i,\rho_i\}$ of states in $\,\S(\H_A)$ with the
full rank average state such that
$\;\mathrm{sup}_i\,\mathrm{rank}\rho_i<\dim \H_E\,$ and
$\,\chi(\{\pi_i,\rho_i\})<+\infty$.}\smallskip

B) \emph{$\chi(\Phi(\mu))<\chi(\mu)$ for any probability
measure $\mu$ on $\,\S(\H_A)$ with the full rank barycenter
such that
$\;\mathrm{sup}_{\rho\in\mathrm{supp}\mu}\,\mathrm{rank}\rho<\dim
\H_E\,$ and $\chi(\mu)<+\infty$.}

\end{property}\medskip

\begin{remark}\label{full-rank-n} By the Stinespring representation every quantum channel is isomorphic to a
particular subchannel of a partial trace. Since the Holevo
quantity does not strict decrease for all channels, Proposition
\ref{main-c-4} clarifies necessity of the full rank average state
condition in Corollary \ref{main-th-c} and in Theorem \ref{main++}. $\square$
\end{remark}\medskip

The conditional entropy of a state $\rho$ of a composite system $AB$
is defined as follows
$$
H_{A|B}(\rho)\doteq H(\rho)-H(\Tr_{\H_A}\rho)
$$
provided
\begin{equation}\label{f-cond}
H(\rho)<+\infty\quad \textrm{and}\quad H(\Tr_{\H_A}\rho)<+\infty.
\end{equation}

By Remark \ref{ent-g-c} concavity of the function $\rho\mapsto
H_{A|B}(\rho)$ on the convex set defined by condition (\ref{f-cond})
follows from monotonicity of the Holevo quantity. Proposition
\ref{main-c-4}A implies the following strict concavity property of
the conditional entropy.\medskip

\begin{corollary}\label{main-c-5} \emph{Let $\rho$ be a full rank state in $\,\S(\H_{A}\otimes\H_{B})$ satisfying (\ref{f-cond}).
Then}
$$
H_{A|B}(\rho)>\sum_i \pi_i H_{A|B}(\rho_i)
$$
\emph{for any ensemble $\{\pi_i,\rho_i\}$ with the average state
$\rho$ such that $\,\mathrm{rank}\rho_i<\dim\H_A$ for all $\,i$.}
\end{corollary}\medskip

By using Proposition \ref{main-c-4}B one can obtain a continuous
(integral) version of Corollary \ref{main-c-5}.

It is easy to construct an example showing that the  strict
concavity property of the conditional entropy stated in Corollary
\ref{main-c-5} does not hold for arbitrary state $\rho$ and its convex decomposition.

\subsubsection{A necessary condition for the equality $\bar{C}(\Phi,\rho)=I(\Phi,\rho)$}

The constrained Holevo capacity $\bar{C}(\Phi,\rho)$ and the quantum
mutual information $I(\Phi,\rho)$  are important
entropic characteristics playing the basic roles in expressions for
the classical capacity and the classical entanglement-assisted
capacity of (constrained or unconstrained) quantum channel $\Phi$ \cite{BSST+,H-c-w-c,N&Ch}.
In general
\begin{equation}\label{n-b-ineq}
\bar{C}(\Phi,\rho)\leq I(\Phi,\rho),\quad \rho\in\S(\H_A)
\end{equation}
(this inequality can be proved by using expression (\ref{mi-rep++})
below valid under the condition $H(\rho)<+\infty$ and a simple approximation).
But there exist channels $\Phi$
for which the equality holds in (\ref{n-b-ineq}) for some states
$\rho$. As the simplest example one can consider the channel
$\Phi(\rho)=\sum_{k}\langle k|\rho|k\rangle|k\rangle\langle k|$,
where $\{|k\rangle\}$ is an orthonormal basis in $\H_B=\H_A$. For
this channel the equality in (\ref{n-b-ineq}) holds for any state
$\rho$ diagonizable in the basis $\{|k\rangle\}$.\medskip

In this subsection we derive from Theorem \ref{main++} a necessary
condition for the equality in (\ref{n-b-ineq}) at some state $\rho$
expressed in terms of the channel $\Phi$.\medskip

The constrained Holevo capacity is defined as
follows
\begin{equation}\label{chi-fun}
\bar{C}(\Phi,\rho)\doteq\sup_{\{\pi_i,\rho_i\},\,\bar{\rho}=\rho}\chi(\{\pi_i,\Phi(\rho_i)\})=\sup_{\mu\in\P(\S(\H_A)),\,
\bar{\rho}(\mu)=\rho}\chi(\Phi(\mu)),
\end{equation}
where the second expression can be derived from Corollary 1 in
\cite{H-Sh-2} with $A=\{\rho\}$.\footnote{In \cite{H-Sh-2} the
constrained Holevo capacity $\bar{C}(\Phi,\rho)$ is denoted
$\chi_{\Phi}(\rho)$ and called the
$\chi$\nobreakdash-\hspace{0pt}function of the channel $\Phi$. We do
not use this notation, since in this paper the symbol $\chi$ denotes
the Holevo quantity of an ensemble of quantum states.} If
$H(\Phi(\rho))<+\infty$ then
\begin{equation*}
\bar{C}(\Phi,\rho)=H(\Phi(\rho))-\hat{H}_{\Phi}(\rho),
\end{equation*}
where $\hat{H}_{\Phi}(\rho)=\inf_{\{\pi_i,\rho_i\},\,\bar{\rho}=\rho}\sum_i\pi_i
H(\Phi(\rho_i))$ (the infimum here can be taken over ensembles of pure states by concavity of the function $\rho\mapsto
H(\Phi(\rho))$).\medskip

In finite dimensions the quantum mutual information is defined as
follows (cf.\cite{N&Ch})
\begin{equation}\label{mi-rep+}
I(\Phi,\rho)=H(\rho)+H(\Phi(\rho))-H(\widehat{\Phi}(\rho)).
\end{equation}
Since in infinite dimensions the terms in the right side of
(\ref{mi-rep+}) may be infinite, it is reasonable to define the
quantum mutual information by the following formula
$$
I(\Phi,\rho) = H\left(\Phi \otimes \id_{R}
(|\varphi_{\rho}\rangle\langle\varphi_{\rho}|)\, \|\, \Phi (\rho)
\otimes \varrho\right),
$$
where $\varphi_{\rho}$ is a purification vector\footnote{This means
that $\Tr_{\H_R}|\varphi_{\rho}\rangle\langle\varphi_{\rho}|=\rho$.}
in $\H_A\otimes\H_R$ for the state $\rho\in\S(\H_A)$ and
$\varrho=\Tr_{\H_A}|\varphi_{\rho}\rangle\langle\varphi_{\rho}|$ is
a state in $\S(\H_R)$ isomorphic to $\rho$. If $H(\rho)$ and
$H(\Phi(\rho))$ are finite then the last formula for $I(\Phi,\rho)$
coincides with (\ref{mi-rep+}).\medskip

\begin{property}\label{inf-dim}
\emph{Let $\,\Phi:\S(\H_A)\rightarrow\S(\H_B)$ be a quantum channel
and $\rho$ be a state in $\,\S(\H_A)$ with the support $\H_{\rho}$
such that $H(\rho)<+\infty$ and the following condition holds
\begin{equation}\label{a-cond}
\exists\;\mu\in\P(\S(\H_A))\;\;\text{such that}\;\;
\bar{\rho}(\mu)=\rho\;\; and\;\; \bar{C}(\Phi,\rho)=\chi(\Phi(\mu)),
\end{equation}
which means that the supremum in the second expression in
(\ref{chi-fun}) is attainable.  If
$\,\bar{C}(\Phi,\rho)=I(\Phi,\rho)<+\infty$ then there exist sets
$\{\varphi_k\}\subset\H_{\rho}$ and $\{\psi_k\}\subset\H_{B}$ such
that
$$
\Phi(\sigma)=\sum_k\langle\varphi_k|\sigma|\varphi_k\rangle|\psi_k\rangle\langle\psi_k|,\qquad
\sum_k|\varphi_k\rangle\langle\varphi_k|=I_{\H_{\rho}},\quad
\|\psi_k\|=1\;\,\forall k,
$$
for any state $\sigma\in\S(\H_{\rho})$ and hence
$\,\Phi|_{\S(\H_{\rho})}$ is an entanglement-breaking
channel.}\smallskip

\emph{Condition (\ref{a-cond}) is valid if either
$H(\Phi(\rho))<+\infty$ or one of the functions $\,\sigma\mapsto
H(\Phi(\sigma)\|\Phi(\rho))$, $\,\sigma\mapsto
H(\widehat{\Phi}(\sigma)\|\widehat{\Phi}(\rho))\,$ is continuous and bounded on the set
$\,\mathrm{extr}\S(\H_A)$.}
\end{property}\medskip

\textbf{Proof.} Without loss of generality we may consider that the measure $\mu$ in
(\ref{a-cond}) belongs to the set $\P(\mathrm{extr}\S(\H_A))$. This follows from convexity of the function
$\sigma\mapsto
H(\Phi(\sigma)\|\Phi(\rho))$, since for an arbitrary measure $\mu\in\P(\S(\H_A))$ there exists a measure $\hat{\mu}\in\P(\mathrm{extr}\S(\H_A))$ such that $\bar{\rho}(\hat{\mu})=\bar{\rho}(\mu)$ and $\int f(\sigma)\hat{\mu}(d\sigma)\geq\int f(\sigma)\mu(d\sigma)$ for any convex lower semicontinuous nonnegative function $f$ on $\S(\H_A)$ (this measure $\hat{\mu}$ can be constructed by using the arguments from the proof of the Theorem in \cite{H-Sh-2}).

By Lemma \ref{inf-dim-l} in the Appendix we have
\begin{equation}\label{mi-rep++}
I(\Phi,\rho)=H(\rho)+\bar{C}(\Phi,\rho)-\bar{C}(\widehat{\Phi},\rho)=\bar{C}(\Phi,\rho)+\Delta_{\Phi}(\rho),
\end{equation}
where
$\Delta_{\Phi}(\rho)=H(\rho)-\bar{C}(\widehat{\Phi},\rho)\geq0$ (by
monotonicity of the Holevo quantity).

Thus $\bar{C}(\Phi,\rho)=I(\Phi,\rho)$ means
$H(\rho)=\bar{C}(\widehat{\Phi},\rho)$. By the remark after Lemma
\ref{inf-dim-l} in the Appendix condition (\ref{a-cond}) implies
that $\bar{C}(\widehat{\Phi},\rho)=\chi(\widehat{\Phi}(\mu))$. Since
$H(\rho)=\chi(\mu)$, equality $H(\rho)=\bar{C}(\widehat{\Phi},\rho)$
shows that the channel $\widehat{\Phi}$ preserves the Holevo
quantity of the measure $\mu$. By Theorem \ref{main++} the
restriction of the channel $\widehat{\widehat{\Phi}}=\Phi$ to the
set $\S(\H_{\rho})$ has the Kraus representation (\ref{Kraus-rep})
such that $\mathrm{rank}V_k=1$ for all $k$.\smallskip

If $H(\Phi(\rho))<+\infty$ then  condition (\ref{a-cond}) holds by
Corollary 2 in \cite{H-Sh-2}.

If the function $\sigma\mapsto H(\Phi(\sigma)\|\Phi(\rho))$ is
continuous and bounded on the set $\,\mathrm{extr}\S(\H_A)$ then the
function $\P(\mathrm{extr}\S(\H_A))\ni\mu\mapsto\chi(\Phi(\mu))$ is
continuous by the definition of the weak convergence. Since the
subset of $\P(\mathrm{extr}\S(\H_A))$ consisting of measures with
the barycenter $\rho$ is compact by Proposition 2 in \cite{H-Sh-2},
the last function attains its least upper bound on this subset.

If the function $\sigma\mapsto H(\widehat{\Phi}(\sigma)\|\widehat{\Phi}(\rho))$ is
continuous and bounded on the set $\,\mathrm{extr}\S(\H_A)$
then the similar arguments shows attainability of the supremum in the definition of the value $\bar{C}(\widehat{\Phi},\rho)$, which is equivalent to (\ref{a-cond}) by the remark after Lemma
\ref{inf-dim-l} in the Appendix. $\square$

\smallskip

We complete this subsection by deriving from Proposition
\ref{inf-dim} a necessary condition for coincidence of the Holevo
capacity with the entanglement-assisted classical capacity of the
channel $\Phi$ with the constraint defined by the inequality
\begin{equation}\label{lc}
 \Tr H\rho\leq h,\quad h>0,
\end{equation}
where $H$ is a positive operator -- Hamiltonian of the input quantum
system.\footnote{Speaking about capacities of infinite dimensional quantum
channels we have to impose particular constraints on the choice of
input code-states to avoid infinite values of the capacities and to
be consistent with the physical implementation of the process of
information transmission \cite{H-c-w-c}.} The operational
definitions of the unassisted and the
entanglement-assisted classical capacities of a quantum channel with
constraint (\ref{lc}) are given in \cite{H-c-w-c}, where the
corresponding generalizations of the HSW and BSST theorems are
proved. \smallskip

The case of unconstrained finite or infinite dimensional channels
can be considered as a partial case of the below observations (by
setting $H=0$).\medskip

The Holevo capacity of the channel $\Phi$ with constraint (\ref{lc})
can be defined as follows
\begin{equation}\label{chi-cap+}
\bar{C}(\Phi|H,h)=\sup_{\Tr H\rho\leq h}\bar{C}(\Phi,\rho).
\end{equation}
By the generalized HSW theorem (\cite[Proposition 3]{H-c-w-c}) the
classical capacity of the channel $\Phi$ with constraint (\ref{lc})
can be expressed by the following regularization formula
$$
C(\Phi|H,h)=\lim_{n\rightarrow+\infty} n^{-1}\bar{C}(\Phi^{\otimes
n}|H_n, nh),
$$
where $H_n=H\otimes I\otimes...\otimes I + I\otimes H\otimes
I\otimes...\otimes I + ... + I\otimes...\otimes I\otimes H$ (each of
$n$ summands consists of $n$ multiples).

By the generalized BSST theorem (\cite[Proposition 4]{H-c-w-c}) the
entanglement-assisted classical capacity of the channel $\Phi$  with
constraint (\ref{lc}) is determined as follows
\begin{equation}\label{ea-cap}
C_{\mathrm{ea}}(\Phi|H,h)=\sup_{\Tr H\rho\leq h}I(\Phi, \rho).
\end{equation}
This expression is proved in \cite{H-c-w-c} under the particular
technical conditions on the channel $\Phi$ and the operator
$H$, which can be removed by using the approximation method \cite{Sh-H}. We will assume that expression (\ref{ea-cap}) is valid.\medskip

Proposition \ref{inf-dim} implies the following necessary condition
for coincidence of $\bar{C}(\Phi|H,h)$ and
$C_{\mathrm{ea}}(\Phi|H,h)$.
\smallskip

\begin{corollary}\label{inf-dim+}
\emph{If $\,\bar{C}(\Phi|H,h)=C_{\mathrm{ea}}(\Phi|H,h)<+\infty$ and
the supremum in (\ref{chi-cap+}) is achieved at a state $\rho_*$
such that $H(\rho_*)<+\infty\,$ and $\,H(\Phi(\rho_*))<+\infty$ then
the restriction of the channel $\,\Phi$ to the set $\,\S(\H_{\rho_*})$,
$\,\H_{\rho_*}=\mathrm{supp}\rho_*$, is entanglement-breaking.}\smallskip

\emph{Instead of the condition $\,H(\Phi(\rho_*))<+\infty$ one can require that one of the functions $\,\sigma\mapsto
H(\Phi(\sigma)\|\Phi(\rho_*))$, $\,\sigma\mapsto
H(\widehat{\Phi}(\sigma)\|\widehat{\Phi}(\rho_*))\,$ is continuous and bounded on the set
$\;\mathrm{extr}\S(\H_A)$.}

\end{corollary}\medskip

\begin{remark}\label{inf-dim++}
If $\Phi$ is an unconstrained finite dimensional channel then the
condition $\bar{C}(\Phi)=\bar{C}(\Phi,\rho_*)$ means that $\rho_*$
is the average state of an optimal ensemble for the channel $\Phi$,
which always exists \cite{Sch-West}. Hence Corollary \ref{inf-dim+}
shows that $\,\bar{C}(\Phi)=C_{\mathrm{ea}}(\Phi)$ implies that
$\Phi$ is an entanglement-breaking channel if there exists an
optimal ensemble for the channel $\Phi$ with the \emph{full rank}
average state. The last condition does not hold in general (see the
example of non-entanglement-breaking channel such that
$\,\bar{C}(\Phi)=C_{\mathrm{ea}}(\Phi)$ considered in \cite{BSST+}).

If $\Phi$ is an infinite dimensional channel then the additional conditions in Corollary \ref{inf-dim+}
imply existence of an optimal measure
$\mu$ for the channel $\Phi$ with constraint (\ref{lc}) such that
$$
\bar{C}(\Phi|H,h)=\int_{\S(\H_A)}H\left(\Phi(\sigma)\Vert
\Phi(\rho_*)\right)\mu(d\sigma),\quad
\bar{\rho}(\mu)=\int_{\S(\H_A)}\sigma\mu(d\sigma)=\rho_*.
$$
These conditions hold if the output entropy of the channel
$\Phi$ (the function $\rho\mapsto H(\Phi(\rho))$) is continuous on
the subset of $\S(\H_A)$ defined by inequality (\ref{lc})
\cite{H-Sh-2}.
\end{remark}
\medskip
\textbf{Example.} The additional conditions in Corollary \ref{inf-dim+} hold for a
Gaussian channel $\Phi$ with the power constraint of the form
(\ref{lc}), where $H=R^{T}\epsilon R$ is the many\nobreakdash-\hspace{0pt}mode oscillator
Hamiltonian (see the remark after Proposition 3 in \cite{H-Sh-2}). In this case the optimal state $\rho_*$ -- the barycenter of an optimal measure -- always exists. So, if we assume that $\rho_*$ is a Gaussian state, then Corollary \ref{inf-dim+} shows that $\bar{C}(\Phi|H,h)=C_{\mathrm{ea}}(\Phi|H,h)$ may be valid only if  $\Phi$ is an entanglement-breaking channel having the Kraus representation with the operators of rank one. \medskip

The above assumption holds provided the conjecture of Gaussian optimizers is valid for the channel $\Phi$ (see \cite{E&W,G-opt} and the references therein).

\section{Appendix}

\subsection{Petz's theorem in infinite dimensions}

Monotonicity of the relative entropy means that
\begin{equation}\label{m-r-e}
    H(\Phi(\rho)\|\Phi(\sigma))\leq H(\rho\|\sigma)
\end{equation}
for any channel $\Phi:\S(\H_A)\rightarrow\S(\H_B)$ and any states
$\rho$ and $\sigma$ in $\S(\H_A)$.

Since finiteness of $H(\rho\|\sigma)$ implies $\mathrm{supp}
\rho\subseteq\mathrm{supp} \sigma$ we will assume in what follows
that $\sigma$ and $\Phi(\sigma)$ are full rank states in $\S(\H_A)$
and in $\S(\H_B)$ correspondingly.

Petz's theorem characterizing the equality case in (\ref{m-r-e}) can
be formulated as follows (where it is assumed that $H(\rho\|\sigma)$ is finite). \smallskip
\begin{theorem}\label{petz-t}
\emph{The equality holds in (\ref{m-r-e}) if and only if
$\,\Theta_{\sigma}(\Phi(\rho))=\rho$, where $\,\Theta_{\sigma}$ is a
channel from $\,\S(\H_B)$ to $\,\S(\H_A)$ defined by the formula}
\begin{equation}\label{theta}
\Theta_{\sigma}(\varrho\,)=[\sigma]^{1/2}\Phi^*\left([\Phi(\sigma)]^{-1/2}(\varrho\,)[\Phi(\sigma)]^{-1/2}\right)[\sigma]^{1/2},
\quad \varrho\in\S(\H_B).
\end{equation}
\end{theorem}
\medskip

Note that $\Theta_{\sigma}(\Phi(\sigma))=\sigma$, so the above
criterion for the equality in (\ref{m-r-e}) can be treated as a
reversibility condition (sufficiency of the channel $\Phi$ with
respect to the states $\rho $ and $\sigma$ in terms of
\cite{P-sqc}).

The proof of (a generalized version of) Theorem \ref{petz-t} in the finite dimensional case can be
found in \cite[Theorem in Sec.5.1]{H&Co}.

In infinite dimensions finiteness of $H(\rho\|\sigma)$ does not
imply that $\lambda\rho\leq\sigma$ for some $\lambda>0$ and hence
the argument of the map $\Phi^*$ in (\ref{theta}) with
$\varrho=\Phi(\rho)$ may be an unbounded operator. Nevertheless, we
can define the channel $\Theta_{\sigma}$ as a predual map to the
linear completely positive unital map
\begin{equation}\label{theta-dual}
\Theta^*_{\sigma}(A)=[\Phi(\sigma)]^{-1/2}\Phi\left([\sigma]^{1/2}A[\sigma]^{1/2}\right)[\Phi(\sigma)]^{-1/2},\quad
A\in\B(\H_A).
\end{equation}
This means that we can use formula (\ref{theta}), keeping in mind
that $\Phi^*$ is an extension of the dual map to unbounded operators
in $\H_B$ (which can be defined by $\Phi^*(\cdot)=\sum_k
V_k^*(\cdot)V_k$ via the Kraus representation $\Phi(\cdot)=\sum_k
V_k(\cdot)V_k^*$).\smallskip

With this definition of the channel $\Theta_{\sigma}$ Theorem
\ref{petz-t} is proved in \cite{P-sqc} (in the von Neumann
algebra settings and with the transition probability instead of the relative entropy)
under the condition that $\rho$ is full rank state
in $\S(\H_A)$. Since in this paper Theorem \ref{petz-t} is used with
the non-full rank state $\rho$, we will show below that it can be
derived from Theorem 3 and Proposition 4 in \cite{J&P}.\footnote{I
would be grateful for any reference on the proof of Theorem
\ref{petz-t} in infinite dimensions without the full rank condition on the state
$\rho$.}\smallskip

Consider the ensemble consisting of two states $\rho$ and $\sigma$
with probabilities $t$ and $1-t$, where $t\in(0,1)$. Let
$\sigma_{t}=t\rho+(1-t)\sigma$. By Donald's identity (Proposition
5.22 in \cite{O&P}) we have
\begin{equation}\label{d-one}
t H(\rho\|\,\sigma)+(1-t)H(\sigma\|\,\sigma)= t
H(\rho\|\,\sigma_{t})+(1-t)H(\sigma\|\,\sigma_{t})+H(\sigma_{t}\|\,\sigma)
\end{equation}
and
\begin{equation}\label{d-two}
\begin{array}{c}
t
H(\Phi(\rho)\|\Phi(\sigma))+(1-t)H(\Phi(\sigma)\|\Phi(\sigma))\\\\=
t
H(\Phi(\rho)\|\Phi(\sigma_{t}))+(1-t)H(\Phi(\sigma)\|\Phi(\sigma_{t}))+H(\Phi(\sigma_{t})\|\Phi(\sigma)),
\end{array}
\end{equation}
where the left-hand sides are finite and coincide by the condition. Since
the first, the second and the third terms in the right-hand side of
(\ref{d-one}) are not less than the corresponding terms in
(\ref{d-two}) by monotonicity of the relative entropy, we obtain
\begin{equation}\label{d-three}
H(\Phi(\rho)\|\Phi(\sigma_{t}))=H(\rho\|\sigma_{t})\quad
\textrm{and}\quad
H(\Phi(\sigma)\|\Phi(\sigma_{t}))=H(\sigma\|\,\sigma_{t}).
\end{equation}
Theorem 3 and Proposition 4 in \cite{J&P} imply
$\rho=\Theta_{t}(\Phi(\rho))$ for all $t\in(0,1)$, where
$$
\Theta_{t}(\varrho\,)=[\sigma_{t}]^{1/2}\Phi^*\left([\Phi(\sigma_{t})]^{-1/2}(\varrho\,)[\Phi(\sigma_{t})]^{-1/2}\right)[\sigma_{t}]^{1/2}, \quad \varrho\in\S(\H_B).
$$

To complete the proof it suffices to show that
\begin{equation}\label{s-lim-one}
\lim_{t\rightarrow +0}\Theta_{t}=\Theta_{\sigma}
\end{equation}
in the strong convergence topology (in which $\Phi_n\rightarrow\Phi$
means $\Phi_n(\rho)\rightarrow\Phi(\rho)$ for all $\rho$
\cite{Sh-H}), since this implies $\rho=\lim_{t\rightarrow
+0}\Theta_{t}(\Phi(\rho))=\Theta_{\sigma}(\Phi(\rho))$.

Since $\Theta_{t}(\Phi(\sigma))=\sigma$ for all $t\in(0,1)$, the set
of channels $\{\Theta_{t}\}_{t\in(0,1)}$ is relatively compact in
the strong convergence topology by Corollary 2 in \cite{Sh-H}. Hence
there exists a sequence $\{t_n\}$ converging to zero such that
\begin{equation}\label{s-lim-two}
\lim_{n\rightarrow+\infty}\Theta_{t_n}=\Theta_0,
\end{equation}
where $\Theta_0$ is a particular channel. We will show that
$\Theta_0=\Theta_{\sigma}$.

Note that (\ref{s-lim-two}) means that the sequence
$\{\Theta^*_{t_n}(A)\}$ tends to the operator $\Theta^*_{0}(A)$ in the weak operator
topology for any positive $A\in\B(\H_B)$.\footnote{Since this topology
coincides with the $\sigma$-weak operator topology on the unit ball
of $\B(\H_A)$ \cite{B&R}.} By Lemma \ref{l-one} below we have
$$
\lim_{n\rightarrow+\infty}[\Phi(\sigma_{t_n})]^{1/2}\Theta^*_{t_n}(A)[\Phi(\sigma_{t_n})]^{1/2}=
[\Phi(\sigma)]^{1/2}\Theta^*_{0}(A)[\Phi(\sigma)]^{1/2}
$$
in the Hilbert-Schmidt norm topology. But the explicit form of
$\Theta^*_{t_n}$ shows that
$$
[\Phi(\sigma_{t_n})]^{1/2}\Theta^*_{t_n}(A)[\Phi(\sigma_{t_n})]^{1/2}=\Phi\left([\sigma_{t_n}]^{1/2}A[\sigma_{t_n}]^{1/2}\right)
$$
and since
$\lim_{n\rightarrow+\infty}[\sigma_{t_n}]^{1/2}A[\sigma_{t_n}]^{1/2}=[\sigma]^{1/2}A[\sigma]^{1/2}$
in the trace norm topology, the above limit coincides with
$\Phi(\left[\sigma]^{1/2}A[\sigma]^{1/2}\right)$. So, we have
$\Theta^*_{0}(A)=\Theta^*_{\sigma}(A)$ for all $A$ and hence
$\Theta_0=\Theta_{\sigma}$.

The above observation shows that for an arbitrary sequence
$\{t_n\}$ converging to zero any partial limit of the sequence
$\{\Theta_{t_n}\}$ coincides with $\Theta_{\sigma}$, which means
(\ref{s-lim-one}).\medskip

\begin{lemma}\label{l-one}
\emph{Let $\,\{\rho_n\}$ be a sequence of states in $\,\S(\H)$
converging to a state $\rho_0$ and $\,\{A_n\}$ be a sequence of
operators in the unit ball of $\,\B(\H)$ converging to an operator $A_0$ in the weak
operator topology. Then the sequence
$\,\{\sqrt{\rho_n}A_n\sqrt{\rho_n}\}$ converges to the operator
$\sqrt{\rho_0}A_0\sqrt{\rho_0}$ in the Hilbert-Schmidt norm topology.}
\end{lemma}

\textbf{Proof.} Since $\{\rho_n\}_{n\geq0}$ is a compact set, the
compactness criterion for subsets of $\S(\H)$ (see \cite[Proposition
in the Appendix]{H-Sh-2}) implies that for an arbitrary
$\varepsilon>0$ there exists a finite rank projector
$P_{\varepsilon}$ such that $\Tr
\bar{P}_{\varepsilon}\rho_n<\varepsilon$ for all $n\geq0$, where
$\bar{P}_{\varepsilon}=I_{\H}-P_{\varepsilon}$. We have
\begin{equation}\label{a-e}
\begin{array}{c}
\sqrt{\rho_n}A_n\sqrt{\rho_n}=\sqrt{\rho_n}P_{\varepsilon}A_n
P_{\varepsilon}\sqrt{\rho_n}\\\\+\sqrt{\rho_n}P_{\varepsilon}A_n
\bar{P}_{\varepsilon}\sqrt{\rho_n}+\sqrt{\rho_n}\bar{P}_{\varepsilon}A_n
P_{\varepsilon}\sqrt{\rho_n}+\sqrt{\rho_n}\bar{P}_{\varepsilon}A_n
\bar{P}_{\varepsilon}\sqrt{\rho_n},\quad n\geq 0,
\end{array}
\end{equation}
Since $P_{\varepsilon}$ has finite rank, $P_{\varepsilon}A_n
P_{\varepsilon}$ tends to $P_{\varepsilon}A_0 P_{\varepsilon}$ in
the norm topology and hence $\sqrt{\rho_n}P_{\varepsilon}A_n
P_{\varepsilon}\sqrt{\rho_n}$ tends to
$\sqrt{\rho_0}P_{\varepsilon}A_0 P_{\varepsilon}\sqrt{\rho_0}$ the
trace norm topology, while it is easy to show that the
Hilbert-Schmidt norm of the other terms in the right-hand side of (\ref{a-e}) tends to zero
as $\,\varepsilon\rightarrow0$ uniformly on $n$. $\square$

\subsection{A difference between the Holevo quantities for complementary channels}

Let $\,\Phi:\S(\H_A)\rightarrow\S(\H_B)$ be a quantum channel and
$\,\widehat{\Phi}:\S(\H_A)\rightarrow\S(\H_E)$ be its complementary
channel. In finite dimensions the \emph{coherent information} of the
channel $\Phi$ at any state $\rho$ can be defined as a difference
between $H(\Phi(\rho))$ and  $H(\widehat{\Phi}(\rho))$ \cite{N&Ch,Sch}.
Since in infinite dimensions these values may be infinite even for
the state $\rho$ with finite entropy, for any such state the
coherent information can be defined via the quantum mutual information as follows
$$
I_c(\Phi,\rho)=I(\Phi,\rho)-H(\rho).
$$

Let $\rho$ be a state in $\,\S(\H_A)$ with finite entropy. By
monotonicity of the Holevo quantity the values $\chi(\Phi(\mu))$ and
$\chi(\widehat{\Phi}(\mu))$ do not exceed $H(\rho)=\chi(\mu)$ for
any measure $\mu\in\P(\mathrm{extr}\S(\H_A))$ with the barycenter
$\rho$. The following lemma can be considered as a
generalized version of the observation in
\cite{Sch}.
\smallskip
\begin{lemma}\label{inf-dim-l}
\emph{Let $\mu$ be a measure in $\P(\mathrm{extr}\S(\H_A))$ with the
barycenter $\rho$. Then
\begin{equation}\label{chi-dif}
\chi(\Phi(\mu))-\chi(\widehat{\Phi}(\mu))=I(\Phi,\rho)-H(\rho)=I_c(\Phi,\rho).
\end{equation}}
\end{lemma}

This lemma shows, in particular, that the difference
$\chi(\Phi(\mu))-\chi(\widehat{\Phi}(\mu))$ does not depend on
$\mu$. So, if the supremum in the second expression in
(\ref{chi-fun}) for the value $\bar{C}(\Phi,\rho)$ is achieved at
some measure $\mu_*$ then the supremum in the similar  expression
for the value $\bar{C}(\widehat{\Phi},\rho)$ is achieved at this
measure $\mu_*$ and vice versa. \smallskip

\textbf{Proof.} If $H(\Phi(\rho))<+\infty$ then
$H(\widehat{\Phi}(\rho))<+\infty$ by the triangle inequality and
(\ref{chi-dif}) can be derived from (\ref{mi-rep+}) by using  the
second formula in (\ref{chi-phi-mu}) and  by noting that the
functions $\rho\mapsto H(\Phi(\rho))$ and $\rho\mapsto
H(\widehat{\Phi}(\rho))$ coincide on the set of pure states. In
general case it is necessary to use the approximation method to
prove (\ref{chi-dif}). To realize this method we have to introduce
some additional notions. \smallskip

Let
$\mathfrak{T}_{1}(\mathcal{H})=\{A\in\mathfrak{T}(\mathcal{H})\,|\,A\geq
0,\;\Tr A\leq 1\}$. We will use the  following two extensions of the
von Neumann entropy to the set $\mathfrak{T}_{1}(\mathcal{H})$
(cf.\cite{L})
$$
S(A)=-\Tr A\log A\quad\textup{and}\quad H(A)=S(A)+\Tr A\log \Tr
A,\quad \forall A \in\mathfrak{T}_{1}(\mathcal{H}).
$$
Nonnegativity, concavity and lower semicontinuity of the von Neumann
entropy imply the same properties of the functions $S$ and $H$ on
the set $\mathfrak{T}_{1}(\mathcal{H})$.

The relative entropy for two operators $A$ and $B$ in
$\mathfrak{T}_{1}(\mathcal{H})$ is defined as follows (cf.\cite{L})
$$
H(A\,\|B)=\sum_{i}\langle i|\,(A\log A-A\log B+B-A)\,|i\rangle,
$$
where $\{|i\rangle\}$ is the orthonormal basis of eigenvectors of
$A$. By means of this extension of the relative entropy the Holevo
quantity of a measure $\mu$ in $\P(\mathfrak{T}_{1}(\H))$ is defined
by expression (\ref{h-q-c}).

A completely positive trace-non-increasing linear map
$\Phi:\mathfrak{T}(\H_A)\rightarrow\mathfrak{T}(\H_B)$ is called
\emph{quantum operation} \cite{N&Ch}.  For any quantum operation $\Phi$ the
Stinespring representation (\ref{Stinespring-rep}) holds, in which
$V$ is a contraction. The complementary operation
$\widehat{\Phi}:\mathfrak{T}(\H_A)\rightarrow\mathfrak{T}(\H_E)$ is defined
via this representation by (\ref{c-channel}).

By the obvious modification of the arguments used in the proof of
Proposition 1 in \cite{H-Sh-2} one can show that the function
$\mu\mapsto\chi(\mu)$ is lower semicontinuous on the set
$\P(\mathfrak{T}_{1}(\H))$ and that for an arbitrary quantum
operation $\Phi$ and a measure $\mu\in\P(\S(\H_A))$ such that
$S(\Phi(\bar{\rho}(\mu)))<+\infty$ the Holevo quantity of the measure
$\Phi(\mu)\doteq\mu\circ\Phi^{-1}\in\P(\mathfrak{T}_{1}(\H_B))$ can
be expressed as follows
\begin{equation}\label{formula}
\chi(\Phi(\mu))=
S(\Phi(\bar{\rho}(\mu))-\int_{\S(\H_A)}S(\Phi(\rho))\mu(d\rho).
\end{equation}

We are now in a position to prove (\ref{chi-dif}) in general case.

Note that for a given measure
$\mu\in\P(\S(\H_A))$ the function
$\Phi\mapsto\chi(\Phi(\mu))$ is lower semicontinuous on the set of
all quantum operations endowed with the strong convergence topology
(in which $\Phi_n\rightarrow\Phi$ means
$\Phi_n(\rho)\rightarrow\Phi(\rho)$ for all $\rho$ \cite{Sh-H}). This follows
from lower semicontinuity of the functional $\mu\mapsto\chi(\mu)$ on
the set $\P(\mathfrak{T}_1(\H_B))$, since for an arbitrary sequence
$\{\Phi_n\}$ of quantum operations strongly converging to a quantum
operation $\Phi$ the sequence
$\{\Phi_n(\mu)\}\subset\P(\mathfrak{T}_1(\H_B))$ weakly converges to
the measure $\Phi(\mu)$ (this can be verified directly by using the
definition of the weak convergence and by noting that for sequences
of quantum operations the strong convergence is equivalent to the
uniform convergence on compact subsets of $\S(\H_A)$).

Let $\{P_n\}$ be an increasing sequence of finite rank projectors in
$\B(\H_B)$ strongly converging to $I_B$. Consider the sequence of
quantum operations $\Phi_n=\Pi_n\circ\Phi$, where
$\Pi_n(\cdot)=P_n(\cdot)P_n$. Then
\begin{equation}\label{c-oper}
    \widehat{\Phi}_n(\rho)=\Tr_{\H_B}P_n\otimes I_{\H_E}V\rho V^*,\quad
    \rho\in\S(\H_A),
\end{equation}
where $V$ is the isometry from Stinespring representation
(\ref{Stinespring-rep}) for the channel $\Phi$.

The sequences $\{\Phi_n\}$ and
$\{\widehat{\Phi}_n\}$ strongly converges to the channels $\Phi$ and
$\widehat{\Phi}$ correspondingly. Let
$\rho=\sum_{k}\lambda_k|k\rangle\langle k|$ and
$|\varphi_{\rho}\rangle=\sum_{k}\sqrt{\lambda_k}|k\rangle\otimes|k\rangle$.
Since $S(\Phi_n(\rho))<+\infty$, the triangle inequality implies
$S(\widehat{\Phi}_n(\rho))<+\infty$. So, we have
\begin{equation}\label{I-n}
\begin{array}{c}
\displaystyle I(\Phi_n,\rho)=H\left(\Phi_n \otimes \id_{R}
(|\varphi_{\rho}\rangle\langle\varphi_{\rho}|)\, \|\, \Phi_n(\rho)
\otimes \varrho\right)\\\\ \displaystyle =
-S(\widehat{\Phi}_n(\rho))+S(\Phi_n(\rho))+a_n=-\chi(\widehat{\Phi}_n(\mu))+\chi(\Phi_n(\mu))+a_n,
\end{array}
\end{equation}
where $a_n=-\sum_{k}\Tr(\Phi_n(|k\rangle\langle
k|))\lambda_k\log\lambda_k$ and the last equality is obtained by
using (\ref{formula}) and coincidence of the functions $\rho\mapsto
S(\Phi(\rho))$ and $\rho\mapsto S(\widehat{\Phi}(\rho))$ on the set
of pure states.

Since the function $\Phi\mapsto I(\Phi,\rho)$ is lower
semicontinuous (by lower semicontinuity of the relative entropy) and
$I(\Phi_n,\rho)\leq I(\Phi,\rho)$  for all $n$ by monotonicity
of the relative entropy under action the quantum operation
$\Pi_n\otimes\id_{\K}$, we have
\begin{equation}\label{lim-r-1}
    \lim_{n\rightarrow+\infty}I(\Phi_n,\rho)=I(\Phi,\rho).
\end{equation}

We will also show that
\begin{equation}\label{lim-r-2}
\lim_{n\rightarrow+\infty} \chi(\Phi_n(\mu))= \chi(\Phi(\mu))\quad
\textrm{and}\quad
\lim_{n\rightarrow+\infty}\chi(\widehat{\Phi}_n(\mu))=
\chi(\widehat{\Phi}(\mu)).
\end{equation}

The first relation in (\ref{lim-r-2}) follows from lower
semicontinuity of the function $\Phi\mapsto\chi(\Phi(\mu))$
(established before) and the inequality $\chi(\Phi_n(\mu))\leq
\chi(\Phi(\mu))$ valid for all $n$ by monotonicity of the Holevo
quantity under action of the quantum operation $\Pi_n$.

To prove the second  relation in (\ref{lim-r-2}) note that
(\ref{c-oper}) implies
$\widehat{\Phi}_n(\rho)\leq\widehat{\Phi}(\rho)$ for any state
$\rho\in \S(\H_A)$. Thus Lemma 2 in \cite{Sh-H} shows that
\begin{equation}\label{chi-ineq}
\chi(\widehat{\Phi}_{n}(\mu))\leq \chi(\widehat{\Phi}(\mu))+
f(\Tr\widehat{\Phi}_{n}(\rho))
\end{equation}
where $f(x)=-2x\log x-(1-x)\log(1-x)$, for any measure
$\mu\in\P(\S(\H_A))$ with finite support and the barycenter $\rho$.
To prove that (\ref{chi-ineq}) holds for any measure
$\mu\in\P(\S(\H_A))$ with the barycenter $\rho$ one can
take the sequence $\{\mu_n\}$ of measures with finite support and
the barycenter $\rho$ constructed in the proof of Lemma 1 in
\cite{H-Sh-2}, which weakly converges to the measure $\mu$, and use
lower semicontinuity of the function $\mu\mapsto\chi(\Psi(\mu))$,
where $\Psi$ is a quantum operation,  and the inequality
$\chi(\widehat{\Phi}(\mu_n))\leq\chi(\widehat{\Phi}(\mu))$ valid for
all $n$ by the construction of the sequence $\{\mu_n\}$ and
convexity of the relative entropy.

Inequality (\ref{chi-ineq}) and lower semicontinuity of the function
$\Phi\mapsto\chi(\Phi(\mu))$ imply the second  relation in
(\ref{lim-r-2}).

Since $\{a_n\}$ obviously tends to $H(\rho)$, (\ref{I-n}),
(\ref{lim-r-1}) and (\ref{lim-r-2}) imply  (\ref{chi-dif}).
$\square$

\vspace{15pt}

I am grateful to A.S.Holevo and to the participants of his seminar "Quantum probability, statistic, information"
(the Steklov Mathematical Institute) for the
useful discussion. I am also grateful to A.Jencova and to T.Shulman for
the valuable help in solving the particular questions.

\end{document}